\documentclass[published]{JHEP3} % 10pt is ignored!

\JHEP{00(2007)000}

\JHEPspecialurl{http://jhep.sissa.it/JOURNAL/JHEP3.tar.gz}

\usepackage{epsfig,multicol,bbm}

\newcommand\fverbdo{\egroup\medskip\noindent%
			\fbox{\unhbox\fverbbox}\ }
\newcommand\fverbit{\egroup\item[\fbox{\unhbox\fverbbox}]}
\newbox\fverbbox

\newcommand{\etal}{ {\it et al.}}

\title{Testing the Copernican and
Cosmological Principles in the local universe with galaxy surveys}

\author{Francesco Sylos Labini \\ Centro Enrico Fermi, Piazza del
 Viminale 1, 0084 Rome Italy \\ and \\ Istituto dei Sistemi Complessi
 CNR, - Via dei Taurini 19, 00185 Rome, Italy\\ E-mail:
 \email{Francesco.SylosLabini@roma1.infn.it}}

\author{Yuri V. Baryshev \\
	Institute of Astronomy, St.Petersburg 
State University, Staryj Peterhoff, 198504,
St.Petersburg, Russia\\ 
	E-mail: \email{yubaryshev@mail.ru}}

\received{\today} 		%%
\accepted{\today}		%% These are for published papers.

\abstract{Cosmological density fields are assumed to be translational
 and rotational invariant, avoiding any special point or direction,
 thus satisfying the Copernican Principle.  A spatially inhomogeneous
 matter distribution can be compatible with the Copernican Principle
 but not with the stronger version of it, the Cosmological Principle
 which requires the additional hypothesis of spatial homogeneity.  We
 establish criteria for testing that a given density field, in a
 finite sample {  at low redshifts}, is statistically and/or
 spatially homogeneous. The basic question to be considered is whether
 a distribution is, at different spatial scales, self-averaging.  This
 can be achieved by studying the probability density function of
 conditional fluctuations.  We find that galaxy structures in the SDSS
 samples, the largest currently available, are spatially inhomogeneous
 but statistically homogeneous and isotropic up to $\sim 100$ Mpc/h.
  Evidences for the breaking of self-averaging are found up to the
 largest scales probed by the SDSS data. The comparison between the
 results obtained in volumes of different size allows us to
 unambiguously conclude that the lack of self-averaging is induced by
 finite-size effects due to long-range correlated fluctuations. 
{  We finally discuss the relevance of these results from the point of
view of cosmological modeling.} 
}


\begin{document} 

%\maketitle  IS IGNORED %%%%%%%%%%%

\section{Introduction}

The attempts to construct cosmological models including spatial
inhomogeneities have experienced a renewed interest in connection with
the evidences for a speeding up expansion of the universe as shown by
the supernovae observations \cite{sn1,sn2}. Indeed, the deduction of
the existence of dark energy is based on the assumption that the
universe has a Friedmann-Robertson-Walker (FRW) geometry. There have
been various claims that these observations can at least in principle
be accounted for without the presence of any dark energy, if we
consider the possibility of inhomogeneities.  This can happen in two
different ways: locally via back-reaction
\cite{buchert,wiltshire,sysky} or by placing the observer in a special
point of the local universe \cite{pedro,celerier}.  Direct
observational tests of the basic assumptions used in the derivation of
the FRW models are thus of considerable importance.  A widespread idea
in cosmology is that the so-called concordance model of the universe
combines {\it two} fundamental assumptions.  The first is that the
dynamics of space-time is determined by Einstein's field
equations. The second is that the universe is homogeneous and
isotropic. This hypothesis, usually called the Cosmological Principle,
is though to be a {\it generalization} of the Copernican Principle
that ``the Earth is not in a central, specially favored position''
\cite{bondi,pedro2}. The FRW model is derived under these two
assumptions and it describes the geometry of the universe in terms of
a single function, the scale factor, which obeys to the Friedmann
equation \cite{weinberg}.  There is a subtlety in the relation between
the Copernican Principle (all observes are equivalent and there are no
special points and directions) and the Cosmological Principle (the
universe is homogeneous and isotropic).  Indeed, the fact that the
universe looks the same, at least in a statistical sense, in all
directions and that all observers are alike does not imply spatial
homogeneity of matter distribution. It is however this latter
condition that allows us to treat, above a certain scale, the density
field as a smooth function, a fundamental hypothesis used in the
derivation of the FRW metric.  Thus there are distributions which
satisfy the Copernican Principle and which do not satisfy the
Cosmological Principle \cite{book}.  These are statistically
homogeneous and isotropic distributions which are also spatially
inhomogeneous.  Therefore the Cosmological Principle represents a
specific case, holding for spatially homogeneous distributions, of the
Copernican Principle which is, instead, much more general.
Statistical and spatial homogeneity refer to two different properties
of a given density field. The problem of whether a fluctuations field
is compatible with the conditions of the absence of special points and
direction can be reformulated in terms of the properties of the
probability density functional (PDF) which generates the stochastic
field. In what follows we precisely discuss this point, both from a
theoretical and observational point of view.

Different strategies have been proposed to test the large scale
isotropy of matter distribution and the basic predictions of
homogeneous models \cite{ellis}.  (i) If the cosmic microwave
background radiation (CMBR) is anisotropic around distant observers,
Sunyaev-Zeldovich scattered photons have a distorted spectrum that
reflects the spatial inhomogeneity \cite{goodman,caldwell}.  (ii)
Tests, based on future supernovae surveys, to determine whether there
is a geometric cusp at the origin \cite{van,pedro}. (iii) Geometric
effects on distance measurements \cite{clarkson2,wiltshire2}. (iv)
There are then some indirect tests \cite{lima}.  All these approaches
thus consider mainly data from the CMBR and from supernovae
surveys and they do not directly test for spatial homogeneity. 

{  However Ellis \cite{ellis2} pointed out that {\it "Spatial
 homogeneity is one of the foundations of standard cosmology, so any
 chance to check those foundations observationally should be welcomed
 with open arms"}.  As recently it became possible to measure directly
 the nature of the spatial galaxy distribution by using galaxy
 redshift surveys, in this paper we present a new test focused to
 determine whether matter distribution is statistically homogeneous
 and isotropic and whether it is spatially homogeneous.} Testing these
 two hypotheses can be achieved by characterizing galaxy distribution
 from the latest data of the Sloan Digital Sky Survey (SDSS)
 \cite{paperdr6}.

{  The paper is organized as follows.  In Sect.\ref{stat} we recall
  some basic statistical properties of spatially homogeneous and
  inhomogeneous distributions. Particularly we discuss that an
  inhomogeneous distribution can be fully compatible with the
  Copernican Principle that there are not special points or
  directions.  The compatibility of a fluctuating density field,
  regardless of whether it is spatially homogeneous, is encoded in the
  properties of its PDF.  This is a very fundamental issue which, in
  our opinion, has been overlooked in the literature. For example, in
  ref. \cite{weinberg} it is stated that {\it "The visible universe
    seems the same in all directions around us, at least if we look
    out to distances larger than about 300 million light years"}, to
  mean that it is spatially homogeneous as then the standard FRW
  modeling is used to derive a number of properties. We point out
  instead that the fact that the observable galaxy distribution looks
  the same in all directions around us implies statistical homogeneity
  and not necessarily spatial homogeneity. This is the key point which
  requires a more detailed investigation from the point of view of
  theoretical modeling, as the lack of spatial homogeneities has a
  deep impact on it.  For instance the works on back-reaction
  \cite{buchert,wiltshire,sysky} consider precisely the effect
  statistically homogeneous large-amplitude fluctuations (i.e.
  spatial inhomogeneities) up to few hundreds Mpc on the geometrical
  properties of the large scale universe.

In Sect.\ref{examples} we present two simple examples which may
clarify this point further.  Note that the discussion in
Sects.\ref{stat}-\ref{examples} refers to an ideal case of a
distribution in an Euclidean space and does not consider the
additional complication introduced by a curved and time dependent
geometry. However this treatment is fully valid in the galaxy samples
we consider, as they are limited to low redshifts, i.e. $z<0.2$. It is
clear that any conclusion we can draw about the statistical properties
of galaxy fluctuations is limited to the range of scales we
considered.

We then pass, in Sect.\ref{data}, to the discussion of the observed
galaxy redshift samples provided by the data release 7 (DR7) of the
Sloan Digital Sky Survey (SDSS). Our main result is that galaxy
distribution as observed by current surveys is inhomogeneous but not
characterized by any special point of direction, i.e. it is
statistically homogeneous but spatially inhomogeneous.  This fact, was
overlooked in the past \cite{pee93} when only the projection on the
sky of the galaxy density field was available.  Having
three-dimensional maps allows us to test statistically homogeneity and
isotropy from many points (observers), which was not possible for
projection on the sky.

In Sect.\ref{discussion} we discuss the relevance of the results
 obtained in the low-redshift galaxy surveys which respect to the
 theoretical modeling and the extension to the test we introduced to
 higher redshift. Particularly we consider the fact that we make
 observations on our past light-cone which is not a space-like
 surface.  Finally we draw our conclusions in
 Sect.\ref{conclusions}. }

\section{Ergodicity and self-averaging}
\label{stat}

Mass density fields can be represented as stationary stochastic
processes.  The stochastic process consists in extracting the value of
the microscopic density function $\rho(\vec{r})$ at any point of the
space. This is completely characterized by its probability density
functional ${\cal P}[\rho(\vec{r})]$. This functional can be
interpreted as the joint probability density function of the random
variables ${\rho}(\vec{r})$ at every point $\vec{r}$.  If the
functional ${\cal P}[\rho(\vec{r})]$ is invariant under spatial
translations then the stochastic process is {\em statistically
  homogeneous} or translational invariant (stationary)
\cite{book}.  When ${\cal P}[\rho(\vec{r})]$ is also invariant under
spatial rotation then the density field is {\em statistically
  isotropic} \cite{book}.
%: the probability distribution on a set of $N$ spatial
%points, depends only on the {\em scalar} distance between all the
%couples of points \cite{book}.

Matter distribution in cosmology then is considered to be a
realization of a {\it stationary} stochastic point process. This is
enough to satisfy the Copernican Principle i.e., that there are no
special points or directions; however this does not imply spatial
homogeneity. Spatially homogeneous stationary stochastic processes
satisfy the special and stronger case of the Copernican Principle
described by Cosmological Principle.  Indeed, isotropy around each
point together with the hypothesis that the matter distribution is a
smooth function of position i.e., that this is analytical, implies
spatial homogeneity. (A formal proof can be found in
\cite{straumann74}.)  This is no longer the case for a non-analytic
structure (i.e., not smooth), for which the obstacle to applying the
FRW solutions has in fact solely to do with the lack of spatial
homogeneity \cite{pwa}.

The condition of {\em spatial homogeneity} ({\em uniformity}) is
satisfied if the ensemble average density of the field $\langle \rho
\rangle$ is strictly positive.  Otherwise, when $\langle \rho
\rangle=0$ the distribution is inhomogeneous. We are interested in the
finite sample properties of a given density field and for this reason
we should introduce the concept of spatial average.
First, we remind that a crucial assumption usually used is that
stochastic fields are required to satisfy spatial {\em ergodicity}.
Let us take a generic observable ${\cal F}= {\cal
  F}(\rho(\vec{r}_1),\rho(\vec{r}_2),...)$ function of the mass
distribution $\rho(\vec{r})$ at different points in space
$\vec{r}_1,\vec{r}_2,...$.  Ergodicity implies that $\left<{\cal
  F}\right> = \overline{{\cal F}} = \lim_{V \rightarrow \infty}
\overline{{\cal F }}_V$, where $\overline{{ \cal F}}_V$ is the spatial
average in a finite volume $V$ \cite{book}.
When considering a finite sample realization of a stochastic process,
and thus statistical estimators of asymptotic quantities, the first
question to be sorted out concerns whether a certain observable is
self-averaging in a given finite volume \cite{Aharony,sdss_aea}.  In
general a stochastic variable $\cal{F}$ is self-averaging if ${\cal F}
= \left< {\cal F}\right> $ (see \cite{sdss_aea} for a more detailed
discussion).
Thus if this is ergodic, $\overline{{\cal F}} = \left< {\cal F}\right>$,
then it is also self-averaging as $\overline{{ \cal F}} = \left<
\overline{{ \cal F} } \right>$: 
%, while if an observable is not
%self-averaging, this does not imply it is not ergodic.  
finite sample spatial averages must be self-averaging in order to
satisfy spatial ergodicity.

A simple test to determine whether a {\it distribution} is stationary
and self-averaging in a given sample of linear size $L$ consists in
studying the probability density function (PDF) of conditional
fluctuations ${ \cal G}$ (which contains, in principle, all
information about moments of any order) in sub-samples of linear size
$L' < L$ placed in different and non-overlapping spatial regions of
the sample (i.e., $S_1,S_2,...S_N$).  That the self-averaging property
holds is shown by the fact that $P({ \cal G},L';S_i)$ is the same,
modulo statistical fluctuations, in the different sub-samples, i.e.,
$ P({ \cal G},L';S_i) \approx P({ \cal G},L';S_j) \; \forall i \ne j.
$
%
%If this is the case then the different statistical quantities may
%satisfy the requirement to be self-averaging.  
On the other hand, if determinations of $P({ \cal G},L'; S_i)$ in
different sample regions $S_i$ show {\it systematic} differences, then
there are two different possibilities: (i) the lack of the property of
stationarity or (ii) the breaking of the property of self-averaging due
to a finite-size effect related to the presence of long-range
correlated fluctuations.  Therefore while the breaking of statistical
homogeneity and/or isotropy imply the lack of self-averaging property
the reverse is not true. However, if the determinations of the spatial
averages give sample-dependent results, this implies that those
statistical quantities do not represent the asymptotic properties of
the given distribution \cite{sdss_aea}.

 To test statistical and spatial homogeneity it is necessary to employ
 statistical quantities that do not require the assumption of spatial
 homogeneity inside the sample and thus avoid the normalization of
 fluctuations to the estimation of the sample average \cite{sdss_aea}.
 These are  conditional quantities, which describe local
 properties of the distribution. For instance, we consider the number
 of points $N_i(r)$ contained in a sphere of radius $r$ centered on
 the $i^{th}$ point.  This depends on the scale $r$ and on the spatial
 position of the $i^{th}$ sphere's center, namely its radial distance
 $R_i$ from a given origin and its angular coordinates
 $\vec{\alpha}_i$.  Integrating over
 $\vec{\alpha}_i$ for fixed radial distance $R_i$, we obtain that
 $N_i(r)=N(r; R_i)$ \cite {sdss_aea}.

\section{Breaking of self-averaging properties} 
\label{examples}

In order to illustrate an example let us consider a case where
translation invariance is broken. We generate a Poisson-Radial
distribution (PRD) which is a inhomogeneous distribution that can
mimic the effect of a ``local hole'' around the origin. In a sphere of
radius $R_0=1$ we place, for instance, $N=2\cdot 10^5$ points.  In
each bin at radial distance from the sphere center $[R_i,R_{i+1}]$,
and with thickness $\Delta R$, the distribution is Poissonian with a
density varying as $n(R) = n_0 \cdot R \;, $ where $n_0$ is a
constant.
We determine the PDF $P(N;r)$ of conditional fluctuations obtained by
making an histogram of the values of $N(r;R)$ at fixed $r$ (see the
upper panels of Fig.\ref{pdf}).  The whole-sample PDF is clearly
left-skewed: this occurs because the peak of the PDF corresponds to
the most frequent counts which are at large radial distance simply
because shells far-way from the origin contain more points. The spread
of the PDF can easily be related to the difference in the density
between small and large radial distances in the sample.  By computing
the PDF into two non-overlapping sub-samples, nearby to and faraway
from the origin, one may clearly identify the systematic dependence of
this quantity on the specific region where this is measured. This
breaking of the self-averaging properties is caused by the
radial-distance dependence of the  density and thus by the
breaking of translational invariance.

Let us now consider a stationary stochastic distribution, where the
breaking self-averaging properties is due to the effect of large scale
fluctuations. An example is represented by the inhomogeneous toy model
(ITM) constructed as follows. We generate a stochastic point
distribution by randomly placing, in a two-dimensional box of side
$L$, structures represented by rectangular sticks.  We first
distribute randomly $N_s$ points which are the sticks centers: they
are characterized by a mean distance $\Lambda \approx
(L^3/N_s)^{1/3}$.  Then the orientation of each stick is chosen
randomly. The points belonging to each stick are also placed randomly
within the stick area, that for simplicity we take to be $\ell \times
\ell/10$.  The length-scale $\ell$ can vary, for example being
extracted from a given PDF. The number of sticks placed in the box
fixes $\Lambda$.  This distribution is by construction stationary
i.e., there are no special points or directions.  When $\ell \ge L$
and $\Lambda \le L$ but with $\ell$ varying in such a way that there
can been large differences in its size, the resulting distribution is
long-range correlated, spatially inhomogeneous and it can be not
self-averaging.  This latter case occurs when, by measuring the PDF of
conditional fluctuations in different regions of a given sample, one
finds, for large enough $r$, systematic differences in the PDF shape
and peak location (see the bottom panels Fig.\ref{pdf}). These are due
to the strong correlations extending well over the size of the sample.

%\fverb!\FIGURE[pos]{body}!\fverbdo
\smallskip
\FIGURE{\epsfig{file=Fig1.eps,width=15cm}
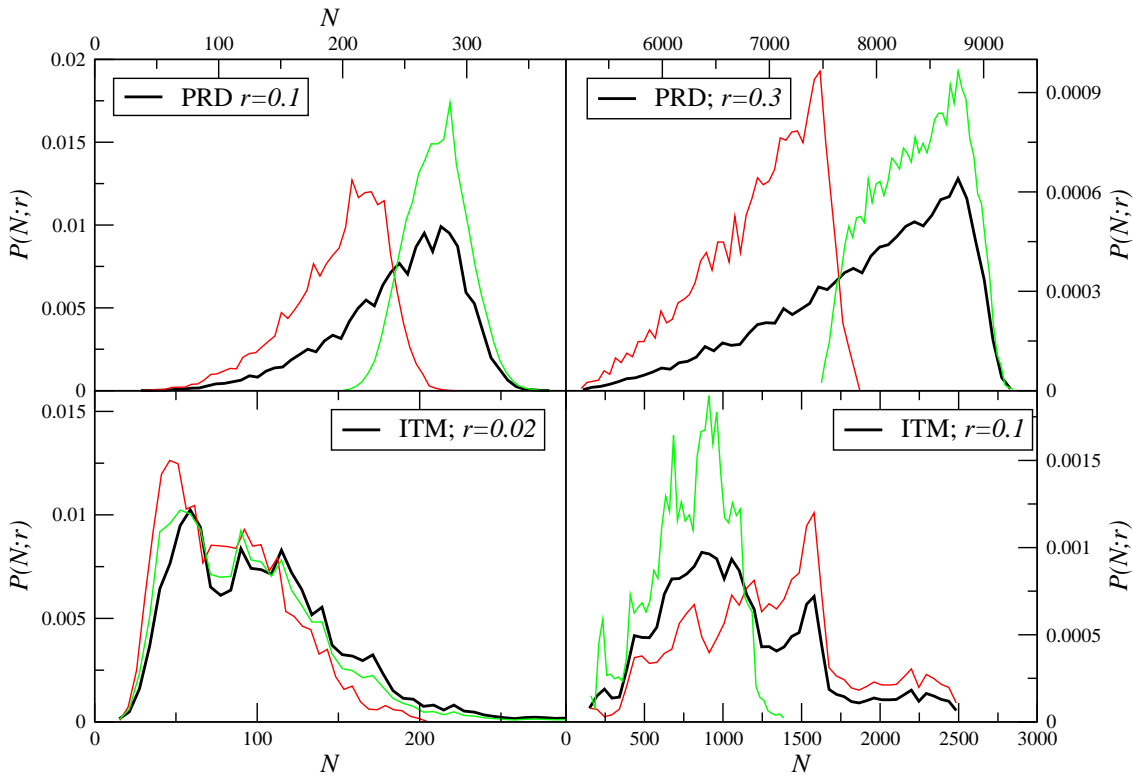
        \caption{Upper panels: The PDF for $r=0.1$ (left) and $r=0.3$
          (right) for PRD computed for the whole sample (black
          line). The red (green) line shows the PDF measured in the
          sub-sample placed closer to (father from) the origin. Lower
          Panels: The same for the ITM at scales $r=0.02$ (left) and
          $r=0.1$ (right)}
	\label{pdf}}

How can we distinguish between the case in which a distribution is not
self-averaging because it is not statistically translational invariant
and when instead this is stationary but fluctuations are too extended
in space and have too large amplitude ?  The clearest test is to
change the scale $r$ where $P(N,r)$ is measured, and determining
whether the PDF is self-averaging. Indeed, in the case of the PRD the
strongest differences between the PDF measured in regions placed at
small and large radial distance from the structure center, occur for
small $r$. This is because the local density has the largest
variations at small and large radial distances by construction.  When
$r$ grows, different radial scales are mixed as the generic sphere of
radius $r$ pick up contributions both from points nearby the origin
and from those far away from it, resulting in a smoothing of local
differences.  Instead, in the ITM for small $r$ the difference is
negligible while for large enough $r$ the different determinations of
the local density start to feel the presence of a few large structures
which dominate the large scale distribution in the sample.

\section{Galaxy Catalogs} 
\label{data}

Let us now consider two (volume limited \cite{sdss_aea}) samples
constructed from the data release 6 (DR6) and DR7 \cite{paperdr6} of
the SDSS (see \cite{sdss_aea,tibor} for details).  We cut each sample
volume into two regions, one nearby us (small $R$~ \footnote{$R=R(z)$
  is the metric distance for which we used the standard cosmological
  parameters $\Omega_M=0.3$ and $\Omega_\Lambda=0.7$. Given that the
  redshift is limited to $z \le 0.2$, different values of
  $\Omega_M,\Omega_\Lambda$ have little effects on our results}) and
the other faraway from us (large $R$). We determine the PDF $P(N;r)$
separately in both regions, and at two different $r$ scales.  In a
first case (left panels of Fig.\ref{vl3}), at small scales ($r=10$
Mpc/h), the distribution is self-averaging both in the DR6 sample
(that covers a solid angle $\Omega_{DR6} =0.94$ sr.)  than in the
sample extracted from DR7 ($\Omega_{DR7}=1.85$ sr. $\approx 2 \times
\Omega_{DR6}$ sr).  Indeed, the PDF is statistically the same in the
two sub-samples considered.  Instead, for larger sphere radii i.e.,
$r=80$ Mpc/h, (right panels of Fig.\ref{vl3}) in the DR6 sample, the
two PDF show clearly a systematic difference. Not only the peaks do
not coincide, but the overall shape of the PDF is not smooth and
different. On the other hand, for the sample extracted from DR7, the
two determinations of the PDF are in very good agreement.  We conclude
therefore that, in DR6 for $r=80$ Mpc/h there are large density
fluctuations which are not self-averaging because of the limited
sample volume \cite{sdss_aea}.  They are instead self-averaging in DR7
because the volume is increased by a factor two.
%

%\fverb!\FIGURE[pos]{body}!\fverbdo
\smallskip
\FIGURE{\epsfig{file=Fig2.eps,width=15cm}
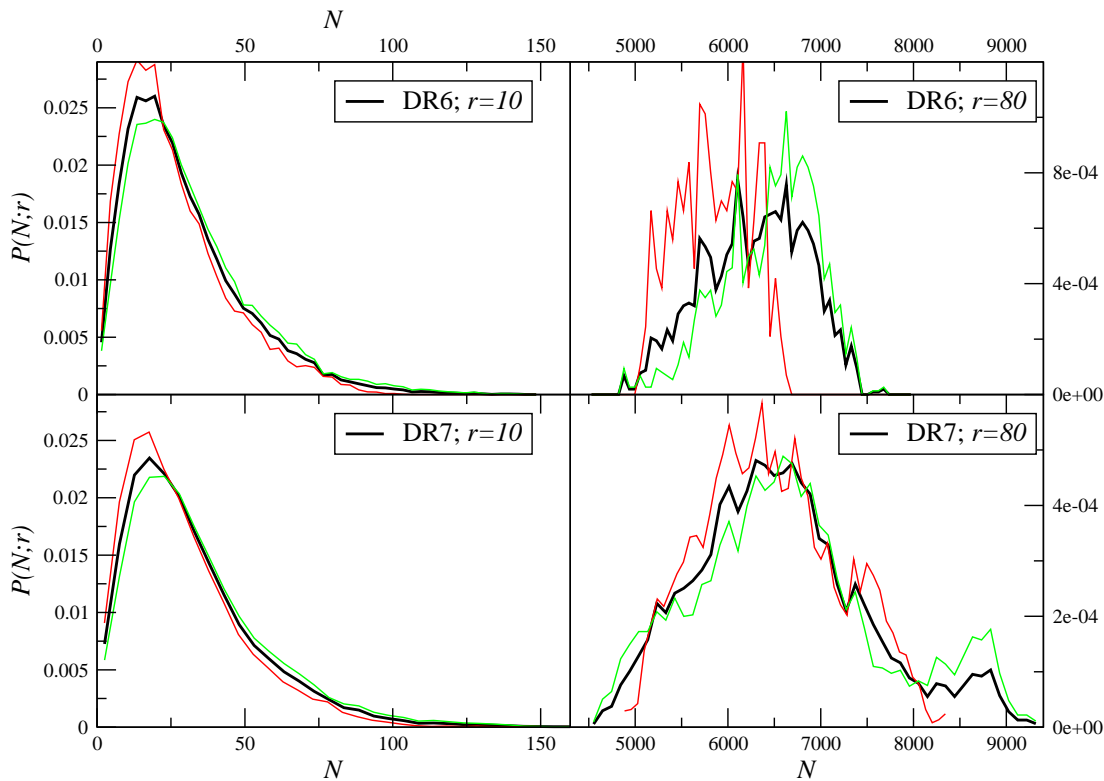
        \caption{PDF of conditional fluctuations in the sample defined by
  $R\in [125,400]$ Mpc/h and $M\in [-20.5,-22.2]$ in the DR6 (upper
  panels) and DR7 (lower panels) data, for two different values of the
  sphere radii $r=10$ Mpc/h and $r=80$ Mpc/h. In each panel, the black
  line represents the full-sample PDF, the red line (green) the PDF
  measured in the half of the sample closer to (farther from) the
  origin.} \label{vl3}}

\smallskip
\FIGURE{\epsfig{file=Fig3.eps,width=15cm}
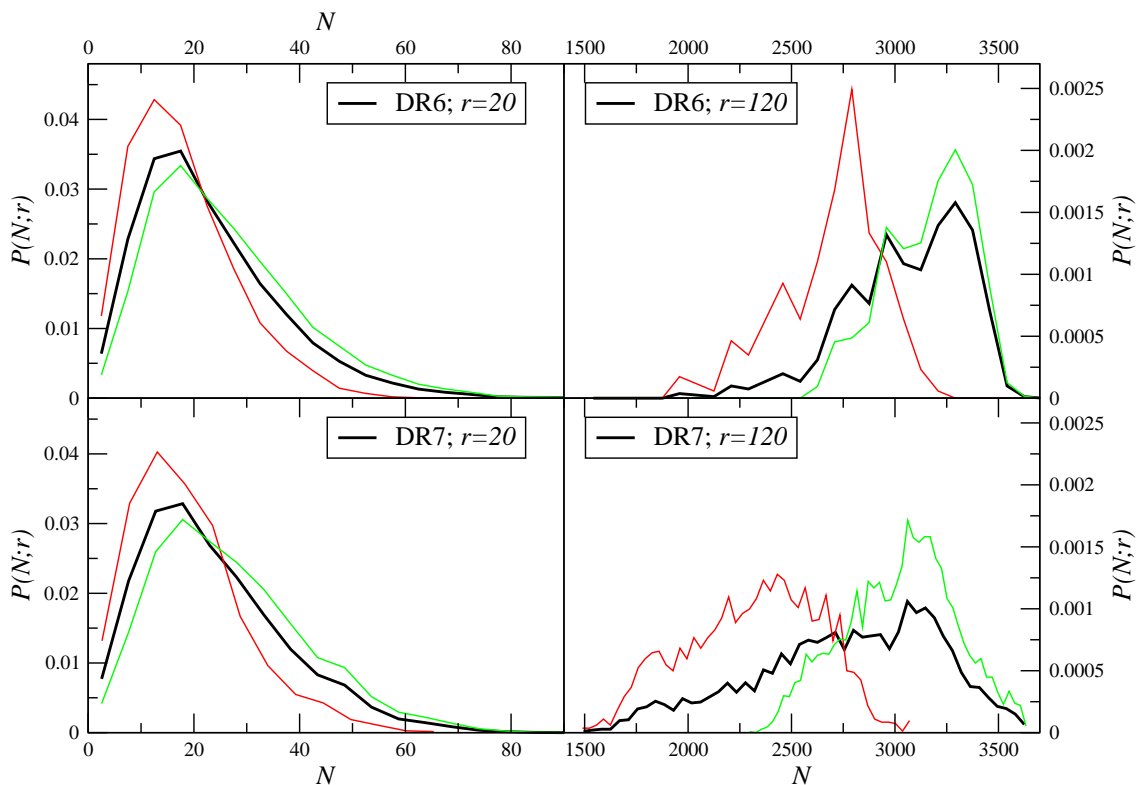
        \caption{The same of Fig.\ref{vl3} but for the sample defined
          by $R\in [200,600]$ Mpc/h and $M\in [-21.6,-22.8]$ and for
          $r=20,120$ Mpc/h.} 	\label{vl5}}

The lack of self-averaging properties at large scales in the DR6
sample is due to the presence of large scale galaxy structures which
correspond to density fluctuations of large amplitude and large
spatial extension, whose size is limited only by the sample
boundaries.  The appearance of self-averaging properties in the larger
DR7 sample volumes is the {\it unambiguous proof} that the lack of
them is induced by finite-size effects due to long-range correlated
fluctuations.

For the deepest sample we consider, which include mainly bright
galaxies, the breaking of self-averaging properties does not occur as
well for small $r$ but it is found for large $r$. This can be due to
the same effects i.e., that the sample volumes are still too small as
even in DR7 for $r =120$ Mpc/h we do not detect self-averaging
properties (right panels of Fig.\ref{vl5}).  Other radial
distance-dependent selections, like galaxy evolution \cite{loveday},
could in principle give an effect in the same direction.  However this
would not affect the conclusion that, on large enough scales,
self-averaging is broken. Note that, contrary to the PRD case, in the
SDSS samples for small values of $r$ the PDF is found to be
statistically stable in different sub-regions of a given sample.  For
this reason we do not interpret the lack of self-averaging properties
as due to a ``local hole'' around us. As discussed above, this would
affect all samples and all scales, which is indeed not the case.
Because of these large fluctuations in the galaxy density field,
self-averaging properties are well-defined only in a limited range of
scales. Only in that range it will be statistically meaningful to
measure whole-sample average quantities \cite{sdss_aea,tibor}.

%%%%%%%%%%%%%%%%%%%%%

\section{Discussion}
\label{discussion}
{  The discussion in the previous sections was meant to treat the
  statistical properties of the galaxy density field in a spatial
  hyper-surface.  As mentioned above, this is an approximation valid
  when considering the galaxy distribution limited to relatively low
  redshifts, i.e. $z<0.2$. In particular, we have developed a test to
  focus on the properties of statistical and homogeneity homogeneity
  in nearby redshift surveys. The assumptions of the cosmological
  model enter in the data analysis when calculating the metric
  distance from the redshift and the absolute magnitude from the
  apparent one and the redshift. However, given that second order
  corrections are small for $z<0.2$, our results are basically
  independent on the chosen underlying model to reconstruct metric
  distances and absolute magnitudes from direct observables. In
  practice we can use just a linear dependence of the metric distance
  on the redshift (which is, to a very good approximation, compatible
  with observations at low redshift). For this same reason we can
  approximate the observed galaxies as lying in a spatial
  hyper-surface.

In the ideal case of having a very deep survey, up to $z\approx 1$, we
should consider that we make observations on our past light-cone which
is not a space-like surface. In order to evolve our observations onto
a spatial surface we would need a cosmological model, which at such
high redshift can play an important role in the whole determination of
statistical quantities.  A sensible question is whether we can to
reformulate the statistical test given so that it can be applied to
data on our past light-cone, and not on an assumed spatial
hyper-surface. Going to higher redshift poses a number of question,
first of the all the one of checking the effect of the assumptions
used to construct metric distances and absolute magnitudes from direct
observables. Testing these effects can be simply achieved by using 
different distance-redshift relations. 

However, we note that a smooth change of the distance-redshift
relation as implied by a given cosmological model, may change the
average behavior of the conditional density as a function of redshift
but it cannot smooth out fluctuations, i.e. it cannot substantially
change the PDF of conditional fluctuations when they are measured
locally. Indeed, our test is based on the characterization of the PDF
of conditional fluctuations and not only of the behavior of the
conditional average density as a function of distance. The PDF
provides, in principle, with a complete characterization of the
fluctuations statistical properties.  We have shown that the PDF of
fluctuations has a clear imprint when the distribution is spherically
symmetric or when it is spatially inhomogeneous but statistically
homogeneous.

 The fact that we analyze conditional fluctuations means that we
 consider only local properties of the fluctuations: local with
 respect to an observer placed at different radial (metric) distances
 from the us, i.e. at different redshifts. For the determination of
 the PDF we have to consider two different length scale: the first is
 the (average) metric distance $R$ of the galaxies on which we center
 the sphere and the second is the sphere radius $r$. Irrespective of
 the value of $R$ when $r$ is smaller than a few hundreds Mpc (i.e.,
 when its size is much smaller than any cosmological length scale), we
 can always locally neglect the specific $R(z)$ relation induced by a
 specific cosmology.  In other words, when the sphere radius is
 limited to a few hundreds Mpc we can approximate the measurements of
 the conditional density to be performed on a spatial hyper-surface.

The whole description of the matter density field in terms of FRW or
even Lemaitre-Tolman-Bondi (LTB) cosmologies, refer to the behavior
of, for instance, the average matter density as a function of time (in
the LTB case also as a function of scale) but it says anything on the
fluctuation properties of the density field.  Thus, when looking at
different epochs in the evolution of the universe, we should detect
that the average density varies (being higher in early epochs). This
means only that the peak of the PDF will be located at different $N$
values, but the shape of the PDF is unchanged by this overall (smooth)
evolution. Fluctuations are simply not present in the FRW or LTB
models, and the whole issue of back-reaction studies is to understand
what is their effect.

Note that models which explain dark energy through inhomogeneity do so
using a spatial under-density in the matter density which varies on
Gpc scales --- out to $z \approx 1$ \cite{pedro}. These models by
placing us at the center of the universe, violate the Copernican
Principle. In this respect we note that, while we cannot make any
claim for $z>0.2$ based on current data, the fact that galaxy
distribution is spatially inhomogeneous but statistically homogeneous
up to 100 Mpc/h, already poses intriguing theoretical
problems. Indeed, in that in that range of scales, the modeling of the
matter density field as a perfect fluid, as required by the FRW
models, is not even a rough approximation. As pointed out by various
authors \cite{slmp98,wiltshire3}, if the linearity of the Hubble law
is a consequence of spatial homogeneity, how is it that observations
show that it is very well linear at the same scales where matter
distribution is inhomogeneous ?  Recently \cite{wiltshire2} it was
speculated a solution to this apparent paradox can be found by
considering both the effects of back-reaction and the synchronization
of clocks. While this is certainly an interesting approach, the
formulation of a more complete and detailed theoretical framework is
still lacking.

Finally we note that there are several complications in radially
 inhomogeneous models at high redshift. Beyond the change of the
 distance-redshift relation, discussed above, another is how structure
 evolves from our past light-cone onto a surface of constant time. Thus
 in order to make a precise test on the spatial properties of a given
 model, one needs to develop the corresponding theory of structure
 formation. However, at least at low redshifts, it seems implausible
 that the main feature of the model, the specific redshift-dependence
 of the spatial density, will not be the clearer prediction for 
the observations of galaxy structures. 
}

%%%%%%%%%%%%%%%%%%%%%

\section{Conclusions}
\label{conclusions} 

We have presented tests on both the Copernican and Cosmological
Principles {  at low redshift, where we can neglect the important
  complications of evolving observations onto a spatial surface for
  which we need a specific cosmological model. We have discussed
  however that the statistical properties of the matter density field
  up to a few hundreds Mpc is crucially important for the theoretical
  modeling.}

We have discussed that these are achieved by considering
the properties of the probability density function of conditional
fluctuations in the available galaxy samples. We have shown that
galaxy distribution in different samples of the SDSS is compatible
with the assumptions that this is transitionally invariant, i.e. it
satisfies the requirement of the Copernican Principle that there are
no spacial points or directions. On the other hand, we found that
there are no clear evidences of spatial homogeneity up to scales of
the order of the samples sizes, i.e. $\sim 100$ Mpc/h
\footnote{{  These results are compatible with those found by
 \cite{2df1,2df2,2df3} in the Two Degree Field Galaxy Redshift
 Survey.}}.  This implies that galaxy distribution is not compatible
 with the stronger assumption of spatial homogeneity, encoded in the
 Cosmological Principle. In addition, at the largest scales probed by
 these samples (i.e., $r\approx 120$ Mpc/h) we found evidences for the
 breaking of self-averaging properties, i.e. that the distribution is
 not statistically homogeneous. Forthcoming redshift surveys will
 allow us to clarify whether on such large scales galaxy distribution
 is still inhomogeneous but statistically stationary, or whether the
 evidences for the breaking of spatial translational invariance found
 in the SDSS samples were due to selection effects in the data.

\acknowledgments We thank T. Antal and N. L. Vasilyev for fruitful
 collaboration, A.  Gabrielli and M. Joyce for interesting discussions
 and comments.  We also thank T. Clifton, R. Durrer and D. Wiltshire
 for useful remarks. An anonymous referee made a list of interesting
 comments and criticisms which have allowed us to improve the
 presentation of our results.  We acknowledge the use of the Sloan
 Digital Sky Survey data ({\tt http://www.sdss.org}).

\end{document}